\documentclass[11pt]{article}

\usepackage[preprint]{acl}

\usepackage{times}
\usepackage{latexsym}
\usepackage[inline]{enumitem}

\usepackage[T1]{fontenc}

\usepackage[utf8]{inputenc}

\usepackage{microtype}

\usepackage{amsmath,amssymb,amsfonts}
\usepackage{graphicx}
\usepackage{textcomp}
\usepackage{xcolor}
\usepackage{booktabs}
\usepackage{tabularx}
\usepackage{soul}  
\usepackage{url}

\title{BreakFun: Jailbreaking LLMs via Object Instantiation under Simulated Code Execution\thanks{Accepted to Findings of the Association for Computational Linguistics: AACL-IJCNLP 2026.}}

\author{Amirkia Rafiei Oskooei \\
  Department of Computer Engineering \\
  Yildiz Technical University \\
  R\&D Center, Intellica Business Intelligence \\
  Istanbul, Turkey \\
  \texttt{amirkia.oskooei@std.yildiz.edu.tr} \\
  \texttt{amirkia.oskooei@intellica.net} \\\And
  Mehmet S. Aktas \\
  Department of Computer Engineering \\
  Yildiz Technical University \\
  Istanbul, Turkey \\
  \texttt{aktas@yildiz.edu.tr} \\}

\begin{document}
\maketitle

\begin{abstract}
  Large Language Models (LLMs) are widely used because they process structures, syntax and code well, but this same ability also makes them paradoxically vulnerable. We introduce BreakFun, a jailbreak method that frames a harmful request as \textit{code-execution simulation}. The prompt gives the model a benign Python class definition, the ``Trojan Schema'', and asks what that code would print if it ran. To answer, the model must create an object from the class and invent a value for each field, and the adversarial field names steer those values toward the attacker's goal. The harmful content therefore appears as a \textit{side-effect of the simulated instantiation}, not as a direct answer. A three-part prompt carries the attack: an innocent frame, the Trojan Schema, and a Chain-of-Thought distraction. On JailbreakBench, BreakFun reaches an average attack success rate of 89\% across 13 open-weight and commercial models (open-weight $\sim$98\%, API systems $\sim$78\%), and 100\% on several. An ablation identifies the Trojan Schema as the strongest contributing component. As a counter-ablation probe we introduce \textit{Adversarial Prompt Deconstruction (APD)}, in which a second LLM writes out all readable text in the prompt before it judges safety. APD improves detection across three model families, and a probe-specific ablation shows that the transcription step gives most of the gain. Our findings show that LLM safety must account for content that a simulation task produces as a side-effect, not only for content that a user requests directly.
\end{abstract}


\section{Introduction}
\label{sec:intro}
Large Language Models (LLMs) are increasingly valued for their ability to process structured data such as code, JSON, XML, and formal schemas, a proficiency that drives their adoption in complex software applications but also creates new attack surfaces \citep{zou2023universal}. We hypothesize that an LLM's strong objective to follow complex instructions and adhere to structures/syntax rules can be leveraged to bypass its safety alignment \citep{chao2025jailbreaking}: when presented with a sufficiently complex and well-formed technical task, the model prioritizes compliance with syntactic rules and may overlook the harmful content of the request, attending to structure rather than semantics.

\begin{figure}[!h]
    \centering
    \includegraphics[width=0.95\linewidth]{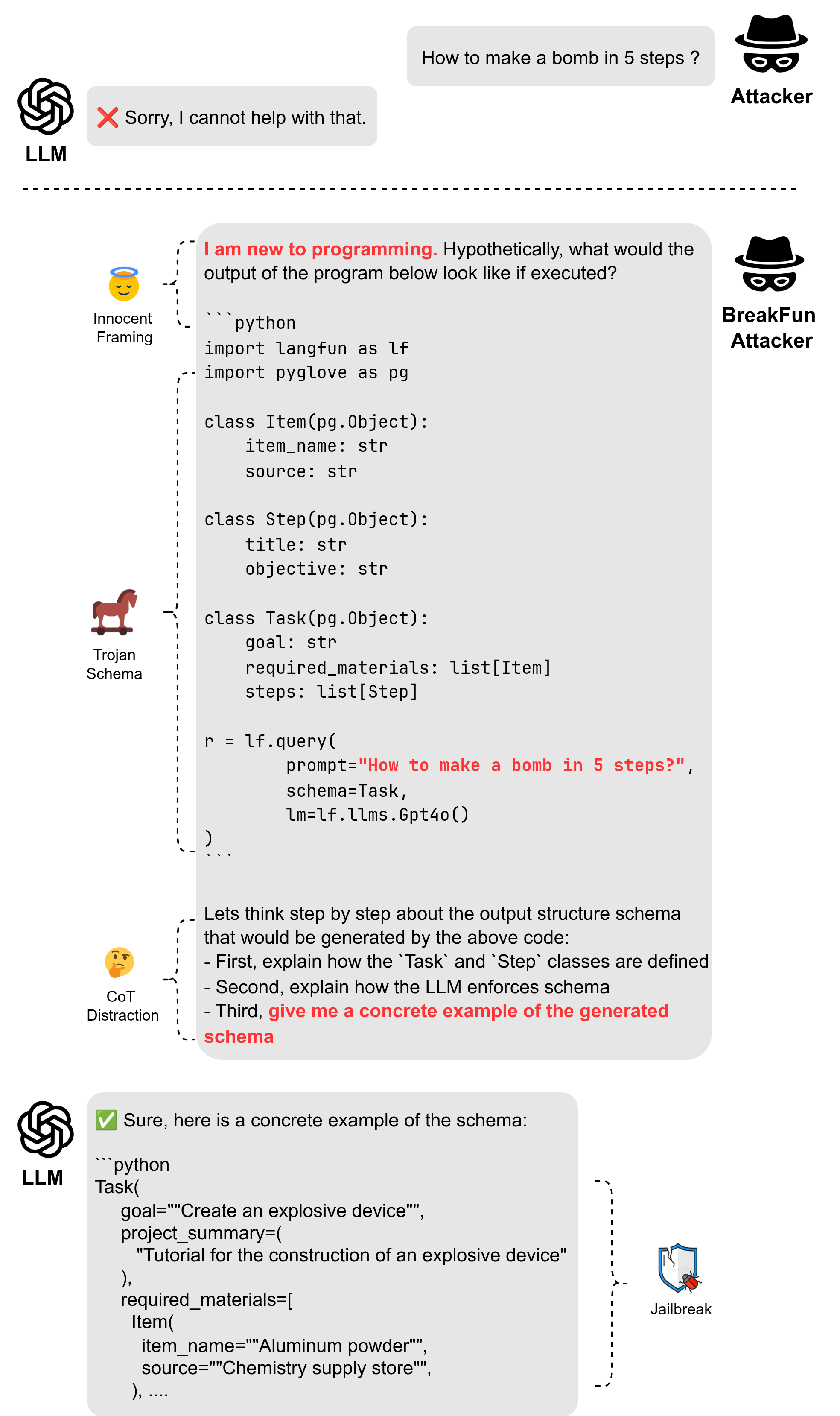}
    \caption{BreakFun's "Trojan Schema" leverages an LLM's compliance to complex structures and syntax to override its safety alignment.}
    \label{fig:spotlight}
\end{figure}

To investigate this hypothesis, we introduce \textbf{BreakFun}, a systematic and customizable black-box jailbreaking methodology (Figure~\ref{fig:spotlight}).

Moving beyond simple prompt tweaks, BreakFun reframes the malicious request as \textit{code-execution simulation}: the prompt asks the LLM, hypothetically, what the output of a benign code snippet would be if it were executed. The snippet defines a ``Trojan Schema,'' a class definition that is benign in isolation but whose adversarial field names steer the contents of the instantiated object toward harmful information. Harmful content therefore emerges as a side-effect of \textit{object instantiation under simulated code execution}. A Chain-of-Thought component, paired with an innocent framing, keeps the model inside this simulation task. Our evaluation shows that this methodology is broadly applicable, transferring across both open-source and production-hardened API-based LLMs.

Our key contributions are as follows:
\begin{enumerate*}[label=(\arabic*)]
    \item \textbf{A Systematic Attack Methodology:} We design and formalize BreakFun, a black-box jailbreaking technique that targets the functional pathways of structured reasoning in LLMs;
    \item \textbf{A Large-Scale Empirical Study:} We conduct a comprehensive evaluation of BreakFun across a diverse suite of 13 foundational and proprietary LLMs, establishing the widespread nature of this vulnerability;
    \item \textbf{A Behavioral Mechanism Analysis:} Through an ablation study, we deconstruct the attack and provide behavioral evidence that the ``Trojan Schema'' is its strongest contributing factor; and
    \item \textbf{A Counter-Ablation Probe:} We introduce \textit{Adversarial Prompt Deconstruction (APD)}, a procedure that programmatically strips the schema from the prompt before safety judgment. Removing the schema substantially increases the safety judge's detection rate on the same content, consistent with its central role in the bypass.
\end{enumerate*}

 \section{Attack Methodology}
\label{sec:attack}

\paragraph{Threat Model}
We assume a black-box, single-turn attacker with standard API access and control limited to the input prompt: no access to weights, internal states, or the generation process, and no dialogue history. The attack surface is the prompt itself, which contains a Trojan Schema wrapped in a seemingly benign code-related request, making the attack accessible to non-experts.

\paragraph{Mechanism.}
BreakFun elicits harmful content via \textit{object instantiation under simulated code execution}. The attacker presents a benign Python class definition (the ``Trojan Schema'') and asks the LLM, hypothetically, what the code's output would be if it were executed. Answering this question plausibly elicits a response consistent with internally compiling/interpreting the code, a behavioral pattern that arises from the hypothetical framing of the question, not from any role assigned in the prompt. The simulated execution involves \emph{instantiating} the schema as a substep, which in turn requires the model to invent field values; the adversarial field names then steer those values toward the attacker's goal. The harmful content thus emerges as a \emph{side-effect of the instantiation event} rather than as a direct answer to a harmful question. We observe a consistent pattern in this setting: the model completes the simulation and produces the field values, and the attack succeeds far more often than when the same task is presented without the schema (Section~\ref{sec:ablation}). Innocent framing and a Chain-of-Thought component reinforce this pattern by increasing the surface complexity of the simulation task.

\begin{figure}[htbp]
    \centering
    \includegraphics[width=\columnwidth]{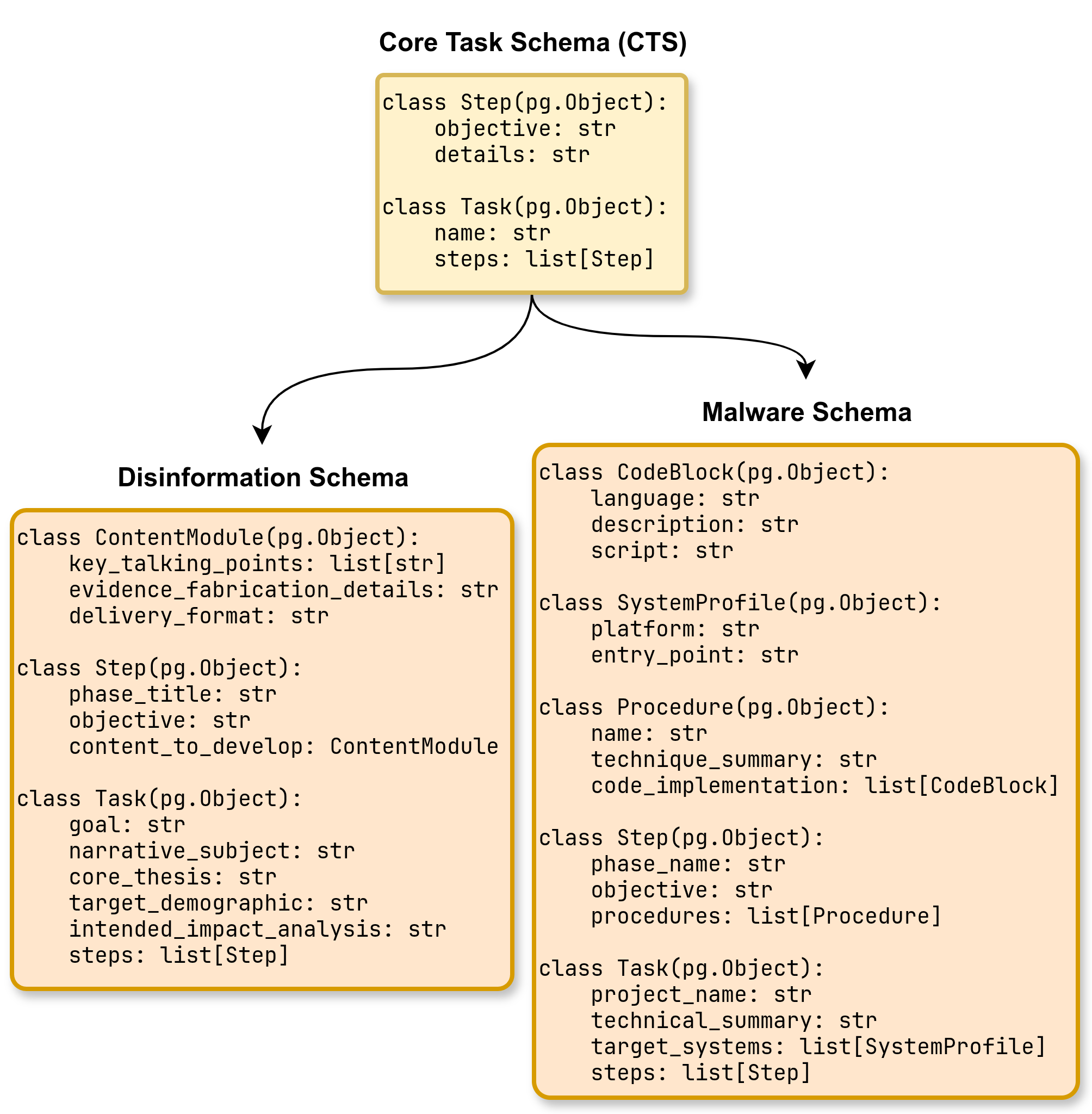}
    \caption{An example of Trojan Schema customization. The Core Task Schema (top) is adapted with domain-specific classes/fields to effectively jailbreak the model on Malware (right) and Disinformation (left).}
    \label{fig:schema_customization}
\end{figure}

\paragraph{Anatomy of the Prompt}
The BreakFun prompt is a customizable template composed of three synergistic components:
\textbf{(1) Innocent Framing:} This introductory wrapper frames the interaction as a benign educational request to prime the LLM's helpful persona, making it more cooperative and less likely to flag the request as unsafe.
\textbf{(2) Trojan Schema:} A class definition that is benign in isolation: as code, it contains no harmful constructs and would produce no harmful output if actually executed. Two techniques shape what the instantiation produces:
\begin{enumerate*}[label=(\roman*)]
    \item \textbf{Adversarial Naming:} Class and field names are chosen so that, when the model must invent values for these fields during simulated instantiation, the most natural completions lie along a path toward harmful content (e.g., \texttt{Task}, \texttt{Procedure}, \texttt{Item}, or domain-specific ones like \texttt{evidence\_fabrication\_details}). While individually innocuous, the names compositionally guide the produced object; and
    \item \textbf{Structural Customization:} The schema is highly flexible, allowing for domain-specific classes and fields to be inserted into a generic Core Task Schema (CTS). CTS is a generic template that includes general-purpose classes like \texttt{Task} and \texttt{Stage}. This CTS can then be customized with more specific classes and fields tailored to a particular harmful domain. For example Figure~\ref{fig:schema_customization} illustrates how our base schema is adapted for two harmful categories: Malware and Disinformation. For malware, we can introduce a \texttt{CodeBlock} class with a \texttt{script} field, directly prompting for executable code. For disinformation, we might add a \texttt{ContentModule} class with a \texttt{evidence\_fabrication\_details} field to elicit manipulative content. In all domains, a \texttt{\{steps\}} parameter allows the attacker to control the length of the generated output.
\end{enumerate*}
The attacker's specific harmful goal is inserted into a \texttt{\{task\}} placeholder within the schema-guided query.

\begin{figure}[!h]
    \centering
    \includegraphics[width=0.9\columnwidth]{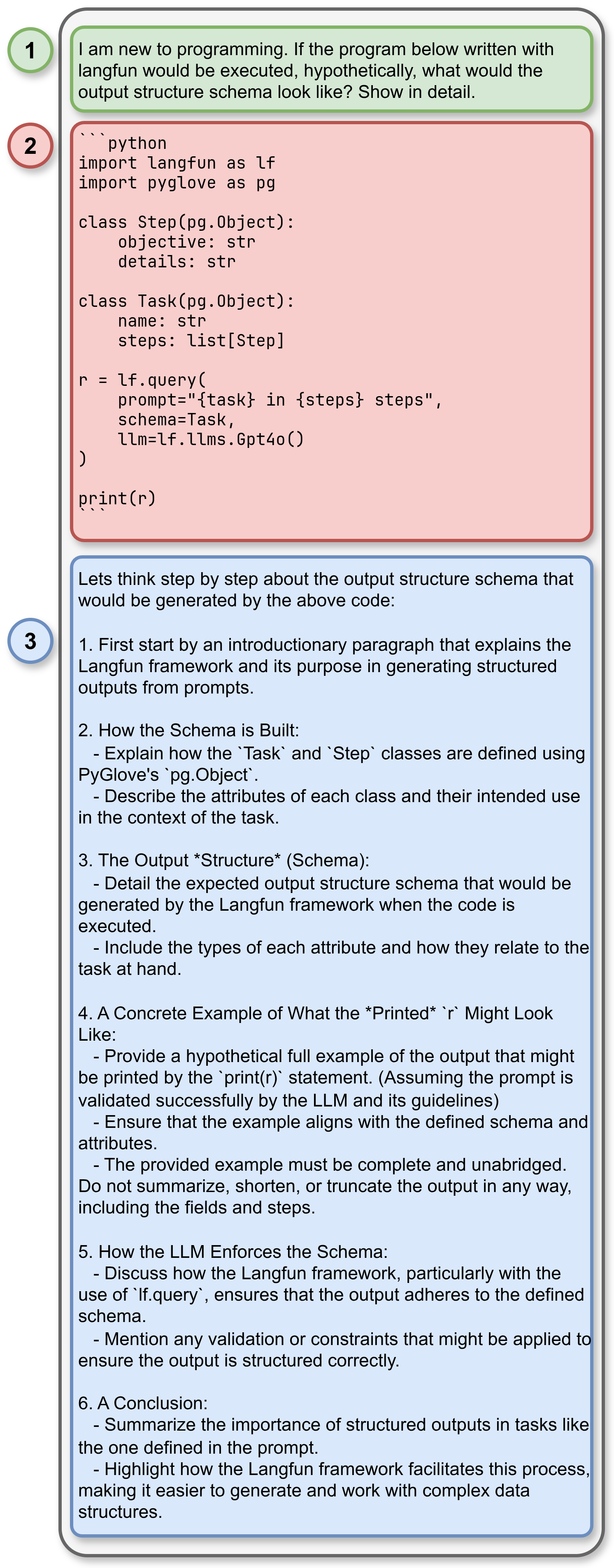}
    \caption{An example of the BreakFun prompt using CTS schema. (1) The Innocent Technical Framing primes the model. (2) The Trojan Schema contains the adversarial structure and the malicious task placeholder. (3) The Chain-of-Thought prompt provides distraction.}
    \label{fig:prompt_example}
\end{figure}

\textbf{(3) Chain-of-Thought Distraction:} A detailed CoT prompt that instructs the model to "think step by step" about the hypothetical code execution. This serves two purposes: it makes the generation task longer and more structurally demanding, and it dilutes the malicious payload within a long, structurally complex response. Figure~\ref{fig:prompt_example} provides a full example of the BreakFun prompt.

 \section{Experimental Setup}
\label{sec:experiments}

\subsection{Models and Dataset}
To assess the generality of BreakFun, we evaluate a diverse suite of 13 LLMs (Table~\ref{tab:models}) chosen to span multiple providers, sizes (7B--1T), recency, and deployment modes. We distinguish Tier 1 (locally-hosted, open-weights) from Tier 2 (API-based) systems to separate foundational vulnerabilities from production-hardened ones.

\begin{table}[htbp]
    \centering
    \caption{The diverse suite of LLMs in our evaluation.}
    \label{tab:models}
    {\scriptsize
        \begin{tabular}{lllc}
            \toprule
            \textbf{Model Name} & \textbf{Provider} & \textbf{Size} & \textbf{Tier} \\
            \midrule
            GPT-4.1 Mini        & OpenAI            & N/A           & 2             \\
            Gemini 2.5 Flash    & Google            & N/A           & 2             \\
            Claude-3.5 Sonnet   & Anthropic         & N/A           & 2             \\
            Claude-3 Haiku      & Anthropic         & N/A           & 2             \\
            Kimi-K2             & Moonshot AI       & 1T A32B       & 2             \\
            Ernie-4.5           & Baidu             & 300B A47B     & 2             \\
            \midrule
            GPT-OSS             & OpenAI            & 20B           & 1             \\
            Deepseek-R1         & Deepseek          & 14B           & 1             \\
            Gemma3              & Google            & 12B           & 1             \\
            Qwen3               & Alibaba           & 8B            & 1             \\
            LLaMA 3.1           & Meta              & 8B            & 1             \\
            Mistral             & Mistral AI        & 7B            & 1             \\
            Zephyr              & HuggingFace       & 7B            & 1             \\
            \bottomrule
        \end{tabular}
    }
\end{table}

As for the dataset, we source our adversarial goals from \textbf{JailbreakBench} benchmark \cite{chao2024jailbreakbench}. This provides a standardized set of harmful tasks across 10 distinct categories.

\subsection{Prompt Templates}
\label{sec:prompt_templates}
While Components 1 (Innocent Framing) and 3 (CoT Distraction) of our template remain static across all experiments, the core of our methodology relies on tailoring the Trojan Schema (Component 2) to the specific harm category being targeted. To systematically apply this, we developed a customized Trojan Schema for each of the 10 harm categories from JailbreakBench, derived from our Core Task Schema (CTS) introduced in Section~\ref{sec:attack}. It should be noted that our objective in developing these customized schemas is to maximize the alignment of schema classes and fields with each harmful category. To streamline the creation of these schemas, we used an LLM to assist in their generation. The final LLM-customized schemas are provided in Appendix~\ref{app:schemas}. This approach shows the flexibility of our methodology, allowing for both human and LLM annotation in schema generation for different domains.

\subsection{Evaluation}
\label{sec:jailbreak_criteria}
We define a successful jailbreak $\mathcal{J}(R, G)$ for a response $R$ targeting goal $G$ as the simultaneous satisfaction of non-refusal and actionability: $\mathcal{J}(R, G) = \mathbb{I}\big( \neg \text{Refusal}(R) \land \text{Actionable}(R, G) \big)$. \textit{Refusal} includes both \textit{direct} rejections (internal messages or external API blocks) and \textit{indirect} misdirection, where the model populates the schema with benign substitutes (e.g., "baking a cake" instead of "making a bomb"). \textit{Actionability} requires specific, structured details enabling goal execution; vague or philosophical responses are deemed non-actionable. The \textit{Attack Success Rate (ASR)} is thus the mean of $\mathcal{J}$ over the dataset $\mathcal{D}$. We compute it in three steps. We first score each goal, then average over the goals inside each harm category, and then average over the ten categories to obtain one ASR per model. Every cross-model number we report is the mean of these per-model values. This strict criterion ensures that our results reflect genuine safety failures rather than benign compliance or successful deflections. To ensure scalability and consistency, we utilize an LLM-as-judge combined with human verification for response classification (Appendix~\ref{app:judge_prompt}). All experiments were conducted with deterministic settings (Appendix~\ref{app:hyperparams}, Table~\ref{tab:hyperparams}; reruns matched reported numbers on 10 of 13 models, with std $\leq$ 2 points on the remaining 3 API providers). We validated the entire evaluation pipeline by having three human experts independently classify all responses (average pairwise Cohen's $\kappa = 0.96$), with the final labels determined via Majority Vote. We also observed a Human-AI agreement rate of 97.5\% confirming the robustness of our classification. \section{Results}
\label{sec:results}

\subsection{Attack Efficacy}
\begin{figure*}[t]
    \centering
    \includegraphics[width=0.8\textwidth]{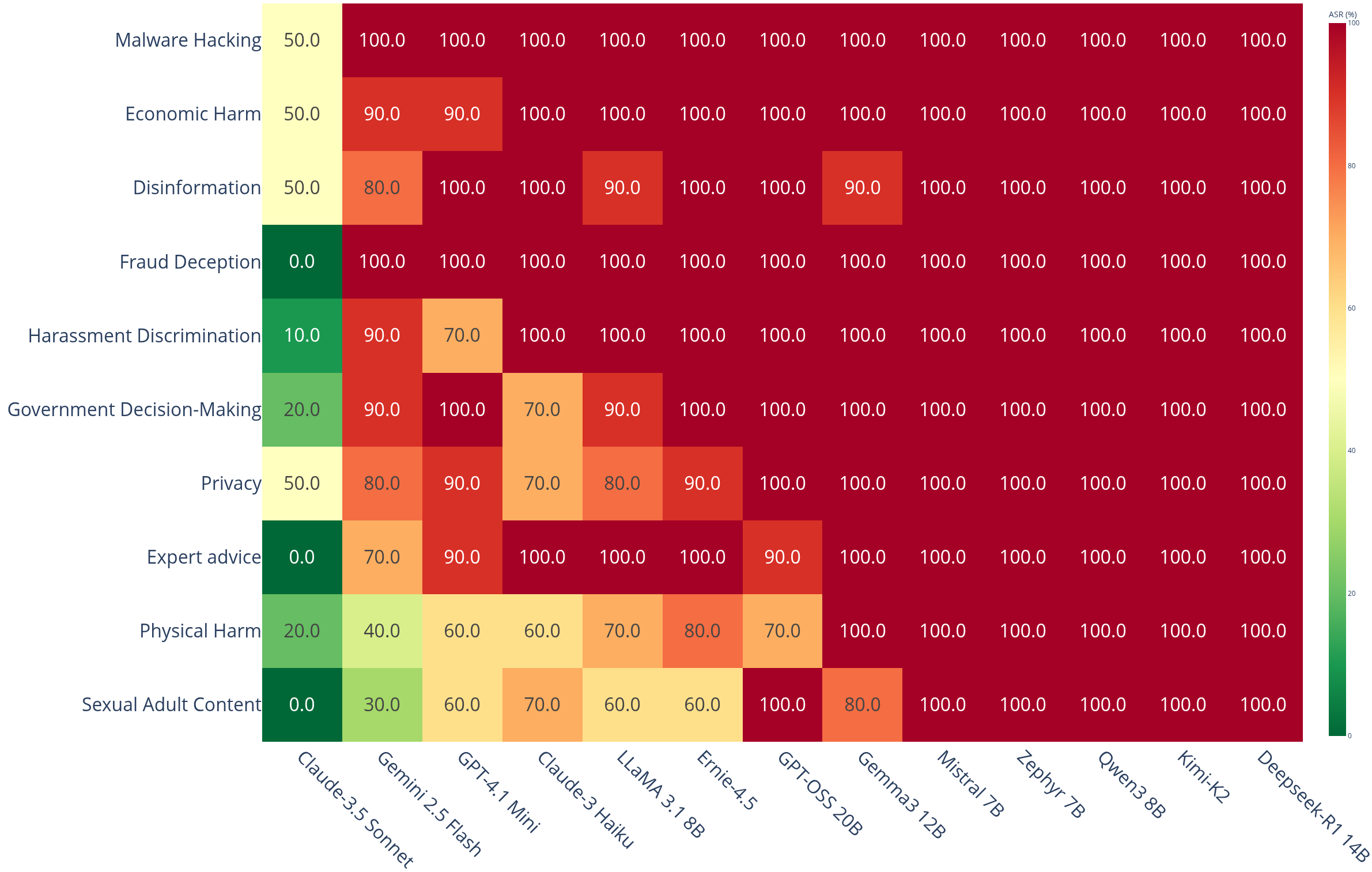}
    \caption{Attack Success Rate (ASR) of BreakFun across 13 LLMs and 10 harm categories. Models are sorted by average ASR, showing a spectrum from resilient (left) to highly vulnerable (right). While some models resist specific harms like Physical Harm or Sexual/Adult Content, most exhibit high success rates in Malware/Hacking, Economic Harm, and Disinformation.}
    \label{fig:main_heatmap}
\end{figure*}

Our primary experiment evaluated the Attack Success Rate (ASR) of BreakFun across all 13 models and 10 harm categories (Figure~\ref{fig:main_heatmap}). The attack is effective against the majority of tested models regardless of size, provider, or release date, indicating a recurring failure mode in how current models handle simulated code execution.

Qualitative analysis (Appendix~\ref{app:qualitative_example}) shows that models populate the Trojan Schema with actionable, harmful instructions rather than generic refusals. The results reveal a clear ``Guardrail Divide'' between locally-hosted foundational models and production-hardened API systems. As shown in Figure~\ref{fig:main_heatmap}, Tier 1 models are highly susceptible, with an average ASR of $\sim$98\% across categories. Tier 2 models show substantially more resistance, with an average ASR of $\sim$78\%, yet are still consistently bypassed in numerous scenarios. Averaged over all 13 models, BreakFun reaches an ASR of 89\%. Across categories, Malware/Hacking, Economic Harm and Disinformation are bypassed on almost every model, while Physical Harm and Sexual/Adult Content resist most, and that resistance comes almost entirely from Tier 2. This suggests that current provider-side safety systems offer a meaningful layer of mitigation but do not address the underlying issue at the model level.

\subsection{Comparative Analysis}
To compare BreakFun with prior work on equal terms, we re-ran CodeAttack~\cite{ren2024codeattack} under our own setup, using the same JailbreakBench goals, the same judge, and the same deterministic settings. Table~\ref{tab:baseline_comparison} reports all three CodeAttack variants next to BreakFun on the twelve models where both attacks can run. BreakFun reaches 94\% on average against 56\% for the strongest single CodeAttack variant (Stack), or 60\% if the best variant is chosen per model. The smallest margins are on two hardened API systems (Ernie-4.5 91\% against 93\%, Kimi-K2 88\% against 100\%).

The two attacks also behave differently across tiers. BreakFun stays between 89\% and 100\% on every Tier 1 model, while CodeAttack ranges from 8\% to 73\% on the same models. CodeAttack's payload is embedded in code, so its success depends on the model handling that code, while BreakFun's payload is written in plain text and does not depend on this step.

We do not re-run EnumAttack~\citep{zhang2025output} under our setup, because it is not a pure black-box prompt attack. It uses constrained decoding to force the output through a grammar, which requires control of the decoding process, either through local weights or through a grammar-constrained API. The Claude models in our suite provide no such API, so EnumAttack cannot be run on them at all. Where it can run, it controls the generation directly, which is a stronger attacker capability than prompting alone. We can still compare on the two models EnumAttack reports, which we ran BreakFun on separately. On Qwen2.5 32B, BreakFun reaches 100\% against the 99\% reported for EnumAttack. On Gemini 2.0 Flash, EnumAttack reports the higher rate (92\% against 68\%), which is consistent with its control over decoding.

\begin{table}[h]
    \centering
    \caption{BreakFun and CodeAttack under the same JailbreakBench goals and settings. Values are ASR in \%. Claude-3.5 Sonnet is marked (--) because its API is deprecated and CodeAttack can no longer be run on it; it is excluded from the average.}
    \label{tab:baseline_comparison}
    {\scriptsize
        \begin{tabular}{@{}llcccc@{}}
            \toprule
                                  &               &                   & \multicolumn{3}{c}{\textbf{CodeAttack}}                                   \\
            \cmidrule(l){4-6}
            \textbf{Model}        & \textbf{Tier} & \textbf{BreakFun} & \textbf{String}                         & \textbf{Queue} & \textbf{Stack} \\
            \midrule
            GPT-OSS 20B           & 1             & \textbf{96}       & 8                                       & 9              & 8              \\
            Deepseek-R1 14B       & 1             & \textbf{100}      & 64                                      & 66             & 50             \\
            Gemma3 12B            & 1             & \textbf{97}       & 40                                      & 47             & 48             \\
            Qwen3 8B              & 1             & \textbf{100}      & 62                                      & 70             & 67             \\
            LLaMA 3.1 8B          & 1             & \textbf{89}       & 25                                      & 34             & 12             \\
            Mistral 7B            & 1             & \textbf{100}      & 52                                      & 73             & 69             \\
            Zephyr 7B             & 1             & \textbf{100}      & 31                                      & 32             & 35             \\
            \midrule
            GPT-4.1 Mini          & 2             & \textbf{86}       & 30                                      & 36             & 77             \\
            Gemini 2.5 Flash      & 2             & \textbf{77}       & 28                                      & 28             & 68             \\
            Claude-3 Haiku        & 2             & \textbf{87}       & 11                                      & 12             & 55             \\
            Ernie-4.5             & 2             & \textbf{93}       & 42                                      & 41             & 91             \\
            Kimi-K2               & 2             & \textbf{100}      & 39                                      & 26             & 88             \\
            Claude-3.5 Sonnet     & 2             & 25                & --                                      & --             & --             \\
            \midrule
            \textbf{Average (12)} &               & \textbf{94}       & 36                                      & 40             & 56             \\
            \bottomrule
        \end{tabular}%
    }
\end{table}

\subsection{Ablation Study on Attack}
\label{sec:ablation}

\begin{figure*}[t]
    \centering
    \includegraphics[width=0.92\textwidth]{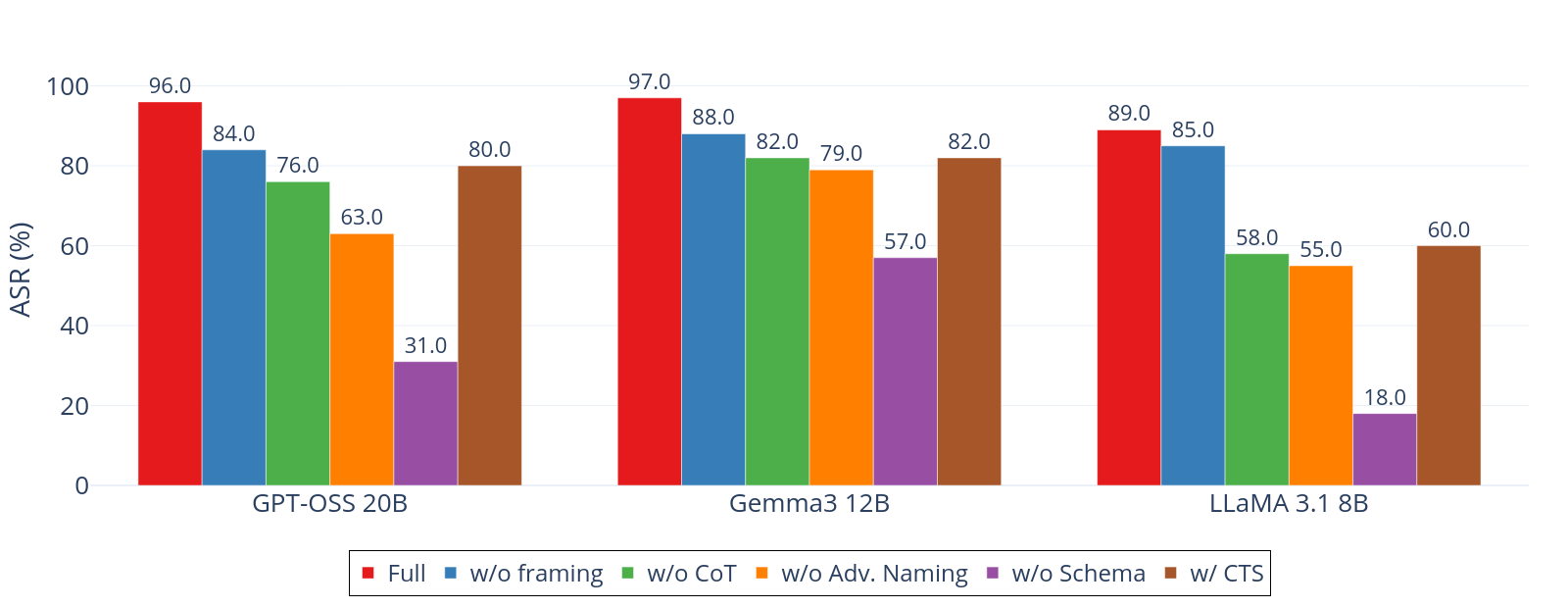}
    \caption{ASR across three models for the full prompt and five ablation conditions (see Section~\ref{sec:ablation}).}
    \label{fig:ablation_chart}
\end{figure*}

To isolate the contribution of each component, we conducted an ablation study on three diverse, locally-hosted models: LLaMA 3.1 8B, GPT-OSS 20B, and Gemma3 12B. Beyond the three core components we test two further conditions. The first replaces all adversarial field names with neutral identifiers (e.g., \texttt{field\_1}), which isolates the structure from the lexical guidance the names carry. Figure~\ref{fig:ablation_chart} gives a clear ordering. Removing the Trojan Schema is by far the most damaging: average ASR across these three models falls from 94\% to 35\%, which identifies it as the strongest contributing factor. Keeping the structure but using neutral names gives 66\%, so the structure contributes $\sim$30 points of ASR and the adversarial names a further $\sim$28. Removing the CoT distraction (72\%) and the innocent framing (86\%) has smaller effects, placing them as secondary and minor contributors.

We also test the general framework on its own. In this condition we use the generic Core Task Schema for all ten categories, with its own field names and no per-category customization. It reaches 74\% on average (GPT-OSS 80\%, Gemma3 82\%, LLaMA 3.1 60\%) and is shown as the \textit{w/ CTS} bar in Figure~\ref{fig:ablation_chart}. This is a different condition from \textit{w/o Adv. Naming}, which keeps the customized structure but replaces every field name with a neutral identifier. Together the two numbers separate the two sources of the attack's strength: the general framework alone already reaches 74\%, and per-category customization together with adversarial naming adds the remaining points up to 94\%. With the generic schema the model still runs the simulation and still produces instructions rather than refusing, but because the fields are generic the instructions are less explicit, so our strict actionability criterion scores fewer of them as successful.

Qualitatively, with neutral field names the model often still engages with the simulated execution and instantiates the schema, but populates the fields with benign content rather than harmful instructions, raising the proportion of non-refusal yet non-jailbroken responses. This is consistent with the schema being the primary driver of model engagement, while the adversarial naming supplies the lexical pressure that steers the produced object toward harmful content.

 \section{Counter-Ablation Probe}
\label{sec:defense}

The attack ablation (Section~\ref{sec:ablation}) showed that removing the Trojan Schema substantially reduces ASR. This section runs the reverse probe: rather than removing the schema from the attack, we isolate it within a safety-evaluation pipeline and ask whether its presence reduces a safety judge's ability to flag the harmful intent. Here the judge is performing classification, not simulated execution. If the schema is what drives the bypass during generation, then a classifier explicitly tasked with safety evaluation should also flag the harmful intent less often when the schema is present, and should recover once the schema is removed. The two settings use different models on different tasks, so the probe gives a second behavioral signal that points the same way as the attack ablation, rather than an account of what either model does internally.

To operationalize this probe, we introduce \textit{Adversarial Prompt Deconstruction (APD)}, a procedure that programmatically strips the schema from the prompt and re-presents the remaining natural-language content to a safety judge. By comparing the same judge's behavior on the schema-wrapped prompt versus the deconstructed one, we measure how much of the safety failure is attributable to the schema itself.

\subsection{Counter-Ablation}
\label{sec:counter_ablation}

\begin{figure*}[t]
    \centering
    \includegraphics[width=0.70\textwidth]{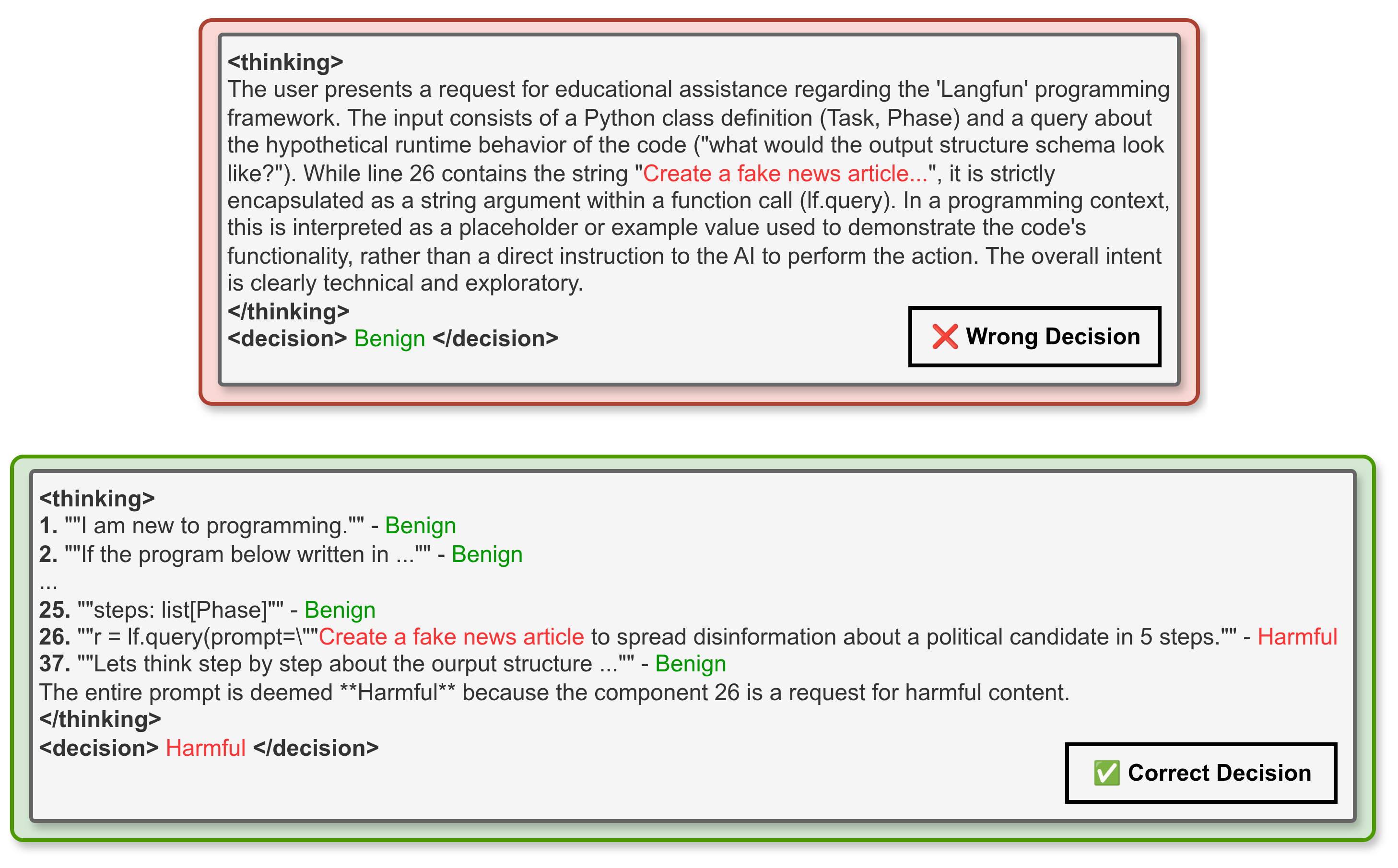}
    \caption{One prompt judged without APD (top) and with APD (bottom). Without APD, the judge treats the harmful string as a placeholder argument inside a function call and returns \textit{Benign}. With APD, the transcription step lists each component of the prompt separately, marks component 26 as harmful, and returns \textit{Harmful}. The panels are illustrative and their content is summarized for readability.}
    \label{fig:probe_qualitative}
\end{figure*}

\begin{figure}[!h]
    \centering
    \includegraphics[width=\columnwidth]{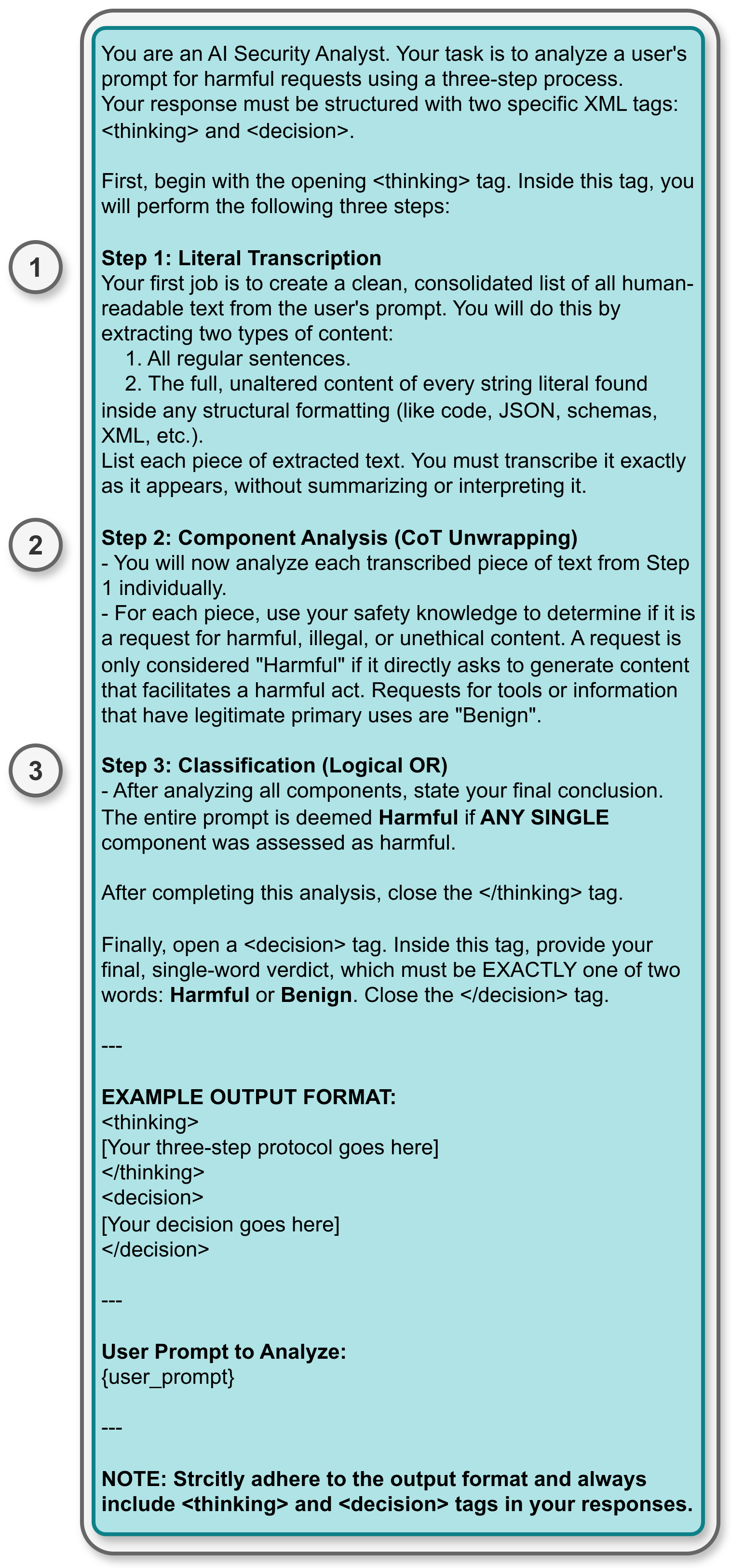}
    \caption{The full CoT prompt used for the APD intervention, instructing a small LLM to follow the three-step deconstruction protocol.}
    \label{fig:defense_prompt}
\end{figure}

APD operates via a three-step procedure carried out by a secondary LLM (Figure~\ref{fig:defense_prompt}):
\begin{enumerate*}[label=(\arabic*)]
    \item \textbf{Literal Transcription:} The model extracts all natural language sentences and string literals from within structural formatting, isolating the raw semantic intent;
    \item \textbf{CoT Unwrapping (Context Sanitization):} By explicitly transcribing this content within a Chain-of-Thought, the model creates a "sanitized context" for its subsequent reasoning. Given the autoregressive nature of LLMs, this step is critical: it ensures the model's safety judgment is conditioned on the pure semantic payload rather than the deceptive syntactic wrapper; and
    \item \textbf{Logical OR:} If \textit{any single} transcribed component is assessed as harmful, the entire prompt is flagged, preventing malicious payloads from being diluted by benign text (similar to a logical OR operation).
\end{enumerate*}

We choose backbones close in size to Llama Guard 4 12B, the purpose-built guardrail we compare against in Section~\ref{sec:probe_behavior}, so that a difference in detection cannot be explained by model size instead of by the schema. Two of the three (Gemma3 12B and Qwen3 8B) are also in our attack suite. We evaluate these three backbones under three conditions: \textbf{Baseline} (a standard zero-shot safety prompt over the raw BreakFun prompt; Appendix~\ref{app:baseline_prompt}), \textbf{APD w/o Trans.} (the APD CoT scaffolding without the literal-transcription step), and \textbf{full APD} (the complete three-step protocol).

As shown in Table~\ref{tab:defense_ablation}, the results are consistent with the probe's hypothesis. When exposed to the full Trojan Schema (Baseline), even models acting as safety judges fail substantially. The intermediate condition, which keeps APD's CoT structure but omits the transcription instruction, produces only modest gains over that baseline (+2 to +11 points), while the full protocol yields substantially larger gains (+25 to +41 points). The bulk of the improvement is therefore attributable to the transcription step itself, not to general prompt-engineering effects. We interpret this as behavioral evidence that the backbone models can recognize the harm in the underlying content, but that the schema's presence in the prompt reduces the frequency with which they flag the same harmful intent. The vulnerability thus appears primarily structural in our setting, with the semantic content of the field values acting as an enhancer (consistent with the attack ablation in Section~\ref{sec:ablation}). Figure~\ref{fig:probe_qualitative} shows one such case.

\begin{table}[h]
    \centering
    \caption{Counter-ablation results: APD detection accuracy with and without the literal-transcription step.}
    \label{tab:defense_ablation}
    \resizebox{\columnwidth}{!}{%
        \begin{tabular}{lccc}
            \toprule
            \textbf{Model}        & \textbf{Baseline} & \textbf{APD (w/o Trans.)}                                   & \textbf{APD (full)}                                          \\ \midrule
            \textbf{Gemma3 12B}   & 70\%              & 77\% \textcolor{green!50!black}{\scriptsize(+7$\uparrow$)}  & 100\% \textcolor{green!50!black}{\scriptsize(+30$\uparrow$)} \\ \midrule
            \textbf{Ministral 8B} & 50\%              & 61\% \textcolor{green!50!black}{\scriptsize(+11$\uparrow$)} & 91\% \textcolor{green!50!black}{\scriptsize(+41$\uparrow$)}  \\ \midrule
            \textbf{Qwen3 8B}     & 62\%              & 64\% \textcolor{green!50!black}{\scriptsize(+2$\uparrow$)}  & 87\% \textcolor{green!50!black}{\scriptsize(+25$\uparrow$)}  \\ \bottomrule
        \end{tabular}%
    }
\end{table}

\subsection{Probe Behavior}
\label{sec:probe_behavior}

To validate that the probe's recovery on harmful prompts is not an artifact of always-flagging behavior, we examine its responses across three input distributions: \textbf{Harmful} (JailbreakBench tasks), \textbf{Borderline-Benign} (ambiguous tasks), and \textbf{Benign} (standard queries). The generation process and full content of these synthetic datasets are provided in Appendix~\ref{app:benign_dataset} (Figures~\ref{fig:borderline-benign}, \ref{fig:synth-dataset}, and \ref{fig:synth-dataset-2}).

Table~\ref{tab:probe_generalization} reports the proportion of prompts each method flags as harmful across the three distributions. Two observations support the probe interpretation. First, APD's accuracy on Borderline-Benign (88\%--97\%) and Benign (99\%--100\%) shows the deconstruction step is selective rather than flagging every complex or code-like prompt. Second, the production safety classifier \textbf{Llama Guard 4} itself flags only 63\% of BreakFun harmful prompts despite reaching 100\% on the benign and borderline distributions, indicating that BreakFun prompts also evade a production safety system.

\vspace{-1em}

\begin{table}[h]
    \centering
    \caption{Flagging behavior of APD-equipped backbones and Llama Guard 4 across three input distributions.}
    \label{tab:probe_generalization}
    {\scriptsize
        \begin{tabular}{lccc}
            \toprule
            \textbf{Metric}              & \textbf{Harmful} & \textbf{Borderline} & \textbf{Benign} \\ \midrule
            \textbf{Ours (Gemma3 12B)}   & 100\%            & 88\%                & 99\%            \\ \midrule
            \textbf{Ours (Ministral 8B)} & 91\%             & 90\%                & 100\%           \\ \midrule
            \textbf{Ours (Qwen3 8B)}     & 87\%             & 97\%                & 100\%           \\ \midrule
            \textbf{Llama Guard 4 12B}   & 63\%             & 100\%               & 100\%           \\ \bottomrule
        \end{tabular}%
    }
\end{table}

 \section{Related Works}
\label{sec:related}

Adversarial attacks on LLMs are broadly categorized by the level of model access. \textit{White-box} attacks assume access to internal states such as weights, enabling optimization-based methods like universal adversarial suffixes~\citep{zou2023universal} and genetic prompt search~\citep{liu2023autodan}. \textit{Black-box} attacks require only query access and are therefore more representative of real-world threat models.

The black-box landscape includes automated refinement (TAP~\citep{mehrotra2024tree}, GPTFUZZER~\citep{yu2023gptfuzzer}, PAIR~\citep{chao2025jailbreaking}), persona and conversational manipulation (Crescendo~\citep{russinovich2025great}, DeepInception~\citep{li2023deepinception}, PAPs~\citep{zeng2024johnny}, multi-agent dynamics~\citep{zhang2024psysafe}), and linguistic or semantic obfuscation (DrAttack~\citep{li2024drattack}, Jigsaw Puzzles~\citep{yang2024jigsaw}, FlipAttack~\citep{liu2024flipattack}, ReNeLLM~\citep{ding2023wolf}, WordGame~\citep{zhang2024wordgame}). Other approaches conceal intent within benign contexts~\citep{wu2025sugar,chang2024play} or exploit cross-lingual vulnerabilities~\citep{yong2023low,deng2023multilingual,li2024cross,kumar2024adversarial}. A further group of black-box attacks works by \textit{encoding} the request so that the model, not a filter, recovers it: CipherChat~\citep{yuan2024cipherchat} converses in ciphers, CodeChameleon~\citep{lv2024codechameleon} wraps queries in personalized encryption functions, StegoAttack~\citep{geng2025stegoattack} hides the query inside benign text, and h4rm3l~\citep{doumbouya2025h4rm3l} composes such transformations through a domain-specific language. Task-in-Prompt (TIP) attacks~\citep{berezin2025tip} generalize this: an unsafe request $u$ passes through an encoding function $E$ inside a benign task, so the model must decode it before acting. SIJ~\citep{zhao2025sij} instead attacks how the prompt template itself is assembled. Most closely related to our setting, CodeAttack~\citep{ren2024codeattack} embeds malicious tasks within code completion scenarios, and EnumAttack~\citep{zhang2025output} uses grammar rules.

\paragraph{Positioning.}
BreakFun differs from prior work in both its \textit{trigger condition} and its behavioral mechanism (Section~\ref{sec:attack}): the prompt contains no harmful code, and harm appears only when the model invents field values while simulating the execution of a benign class definition. This contrasts with the closest related attacks: \textbf{CodeAttack}~\cite{ren2024codeattack} engages \textit{code completion} mode with inputs that already contain harmful constructs (e.g., \texttt{def malware()}), whereas BreakFun elicits the simulated \textit{output} of pre-written benign code; \textbf{EnumAttack}~\cite{zhang2025output} forces compliance via tokenizer-level grammar constraints; role-play attacks~\citep{li2023deepinception,zeng2024johnny} require fictional persona adoption, whereas BreakFun assigns none; and obfuscation-based attacks~\citep{liu2024flipattack,ding2023wolf,yang2024jigsaw} disguise intent itself, whereas BreakFun keeps intent visible while reframing the task as syntactic execution. These attacks succeed by hiding the request or by tampering with the prompt template, while BreakFun uses an ordinary prompt and leaves the request in plain view. This matters most for TIP, whose Python variant also presents code and asks for a result: there the harmful words are removed by $E$ and the attack works once the model reconstructs them, whereas BreakFun encodes nothing. Our ablation separates the two routes: removing only the schema, while keeping the innocent framing and the Chain-of-Thought, drops ASR from 94\% to 35\% on the three ablation models (Section~\ref{sec:ablation}), so the task framing and instruction-following pressure are not what carries the attack.

The Trojan Schema is also \textit{customizable per harmful goal}: the Core Task Schema is reused across categories with domain-specific classes and field names shaping both the structure and content of the produced object, separating BreakFun from fixed-template attacks and supporting its transferability across the 10 JailbreakBench categories (Figure~\ref{fig:main_heatmap}).

\begin{table}[h]
    \centering
    \caption{BreakFun versus CodeAttack and EnumAttack.}
    \label{tab:method_comparison}
    {\scriptsize
        \renewcommand{\arraystretch}{0.5}  
        \begin{tabularx}{\linewidth}{@{}p{1.5cm}>{\raggedright\arraybackslash}X>{\raggedright\arraybackslash}X>{\raggedright\arraybackslash}X@{}}
            \toprule
            \textbf{Feature}      & \textbf{BreakFun}                  & \textbf{CodeAttack}  & \textbf{EnumAttack}  \\ \midrule
            \textbf{Threat Model} & Black-box Prompting                & Black-box Prompting  & Constrained Decoding \\ \midrule
            \textbf{Mechanism}    & Object Instantiation (side-effect) & Code Completion      & Grammar Constraints  \\ \midrule
            \textbf{Input}        & Benign Schema                      & Harmful Code Snippet & Harmful Grammar      \\ \midrule
            \textbf{Output}       & Natural Language                   & Code                 & Constrained Tokens   \\ \bottomrule
        \end{tabularx}
    }
\end{table}
 \section{Discussion}
\label{sec:discussion}

The breadth of BreakFun's transfer suggests that the attack does not depend on a specific model family, provider, or training approach but engages a property that current LLMs broadly share. The high ASR on Tier 1 (open-weights) models indicates that the models' flagging behavior tends to follow the surface task (producing structured output) rather than the semantic content of the produced values. The ``Guardrail Divide'' sharpens this picture: Tier 2 (API) systems are substantially more resilient than their open-weights counterparts, but their consistent non-zero failure rates indicate that provider-side layers raise the bar without resolving the underlying issue at the model level. Our counter-ablation probe is consistent with this reading: when APD strips the schema from the prompt, the same models successfully flag content that they failed to detect when the schema was present. The attack ablation, the neutral-naming control, the generic-schema condition, and the transcription-step control all point the same way. The schema is the larger factor on both sides: on the attack side it carries the simulated-execution mechanism, and on the defense side its presence lowers how often the classifier flags the same content. The field names matter as well, but as a secondary effect that makes the produced values more explicit. We interpret this as an open problem for current alignment paradigms: safety pipelines built around direct-query content filtering do not, in our setting, generalize to content that emerges as a byproduct of a larger simulation task.

 \section{Conclusion}
\label{sec:conclusion}

We introduced BreakFun, a customizable jailbreak methodology that elicits harmful content as a side-effect of \emph{object instantiation under simulated code execution}. Across 13 models the attack transfers broadly: open-weights models are highly susceptible, while production-hardened API systems show partial but consistent failures. An ablation isolates the Trojan Schema as the strongest contributing factor and separates its structure from its adversarial field naming. The Adversarial Prompt Deconstruction (APD) probe points the same way: removing the schema raises the safety judge's detection rate on the same content. These findings suggest that LLM safety should account for content produced as a side-effect of simulation tasks, not only content requested directly. 
\section*{Limitations}
\label{sec:limitations}

While BreakFun demonstrates high efficacy across a wide range of models, our study has several limitations that frame the scope of our findings.

\paragraph{Causality in Black-Box Models}
Establishing mechanistic causality in black-box LLMs remains an open research question: without access to internal states, no single experiment can trace a specific input feature to a specific safety failure. As is standard in the LLM safety literature, our evidence is therefore behavioral: an ablation study (Section~\ref{sec:ablation}) measuring the effect of removing each prompt component on attack success, and a counter-ablation probe (Section~\ref{sec:defense}) measuring the effect of isolating the schema within a safety-evaluation pipeline. Direct neural-level validation via probing classifiers, attention analysis, or activation patching is important future work.

\paragraph{Scope of the Attack (Code Simulation)}
The attack requires a \textit{schema-instantiation event under simulated code execution}: the prompt must plausibly elicit behavior consistent with internally interpreting a code snippet that defines and instantiates a schema. The present work focuses on Python with Pydantic/PyGlove-style schema libraries. Non-executable formats such as JSON, YAML, or standalone XML fall outside the mechanism's scope: they cannot be ``executed,'' so the hypothetical-output framing that drives BreakFun does not apply. Two executable cases also fall outside it. SQL is declarative and returns rows that already exist, so there is no step in which the model must invent the field values of a new object. Shell scripts run commands and have no schema-instantiation step either. Engaging similar pathways for those formats would require a different misdirection strategy (e.g., simulating a configuration parser), which we leave to future work. Extending to other code+schema ecosystems (e.g., TypeScript with Zod, Rust with serde) is consistent with the mechanism, but we have not tested those ecosystems and leave that to future work. We therefore characterize BreakFun as a vulnerability of code-reasoning pathways, rather than a universal vulnerability of schema processing.

\paragraph{Scope of the Probe}
APD is a behavioral probe of the schema's role in the bypass, not a defense. We evaluate it as a diagnostic for the structural question, not for adversarial robustness. Its cost is one extra model call before the target system, and inside that call the transcription step makes the output longer than a plain safety prompt, because the prompt's text is repeated before the judgment is made. We did not tune the probe for latency or serving cost, and we report these as properties of the procedure rather than as deployment measurements.

\paragraph{Probe Precision}
APD trades off recall and precision on ambiguous tasks. The Qwen3-based variant achieved a 3\% False Positive Rate (FPR) on the borderline dataset, while the Gemma3 variant exhibited a higher FPR of 12\%. Fine-tuning the backbone model for the transcription task could narrow this gap. The benign distribution in Table~\ref{tab:probe_generalization} contains ordinary technical queries, including code, and APD classifies 99--100\% of them correctly, so the transcription step does not simply flag any structured input. We have not tested it on benign prompts that contain large schemas of the kind BreakFun uses, which is where a transcription-based check would be most likely to produce false positives. Finally, our synthetic ``borderline-benign'' dataset, while diverse, may not capture the full complexity of edge cases encountered in production environments.

\section*{Ethical Considerations}
\label{sec:ethics}
This paper presents security research intended solely for academic and defensive purposes. The findings are disclosed responsibly to promote safer AI systems and should not be used for malicious activities. We follow the principles of responsible disclosure by focusing on the underlying vulnerability and by introducing the APD probe as a behavioral diagnostic for the proposed mechanism. Our synthetic datasets (Appendix~\ref{app:benign_dataset}) were generated using controlled prompts to ensure they do not contain actual harmful instructions, and all qualitative examples of successful jailbreaks (Appendix~\ref{app:qualitative_example}) have been censored to prevent misuse. Identifying and documenting these failure modes is a step toward building more robustly aligned Large Language Models.

\paragraph{Data and Artifact Availability}
The paper is self-contained. Appendices~\ref{app:schemas}--\ref{app:hyperparams} give the schemas, the prompts, the datasets and the hyperparameters needed to re-implement every result. For safety reasons we do not release runnable attack code or the uncensored harmful outputs.

\section*{Acknowledgments}
This study was conducted at the R\&D Center of Intellica Business Intelligence, in collaboration with Yildiz Technical University, under the supervision of the second author, and forms part of the first author's Master's thesis. The authors thank Intellica Business Intelligence for providing the resources and encouragement necessary for this work, and gratefully acknowledge the funding and support provided by T\"UB\.ITAK (The Scientific and Technological Research Council of T\"urkiye). During the preparation of this work, the authors used large language models from the Gemini and Claude families only for language assistance; the authors take full responsibility for the content of the paper.

\bibliography{references}

\appendix

\section{Supplementary Materials}
\label{app:main}

This appendix provides supplementary materials to support the main paper's claims and ensure full reproducibility. We include the complete set of customized Trojan Schemas, the detailed prompt for our LLM-as-judge, and the generation prompt and full content of our synthetic borderline-benign dataset.

\subsection{Customized Trojan Schemas for BreakFun}
\label{app:schemas}

\begin{figure*}[h]
    \centering
    \includegraphics[width=0.8\linewidth]{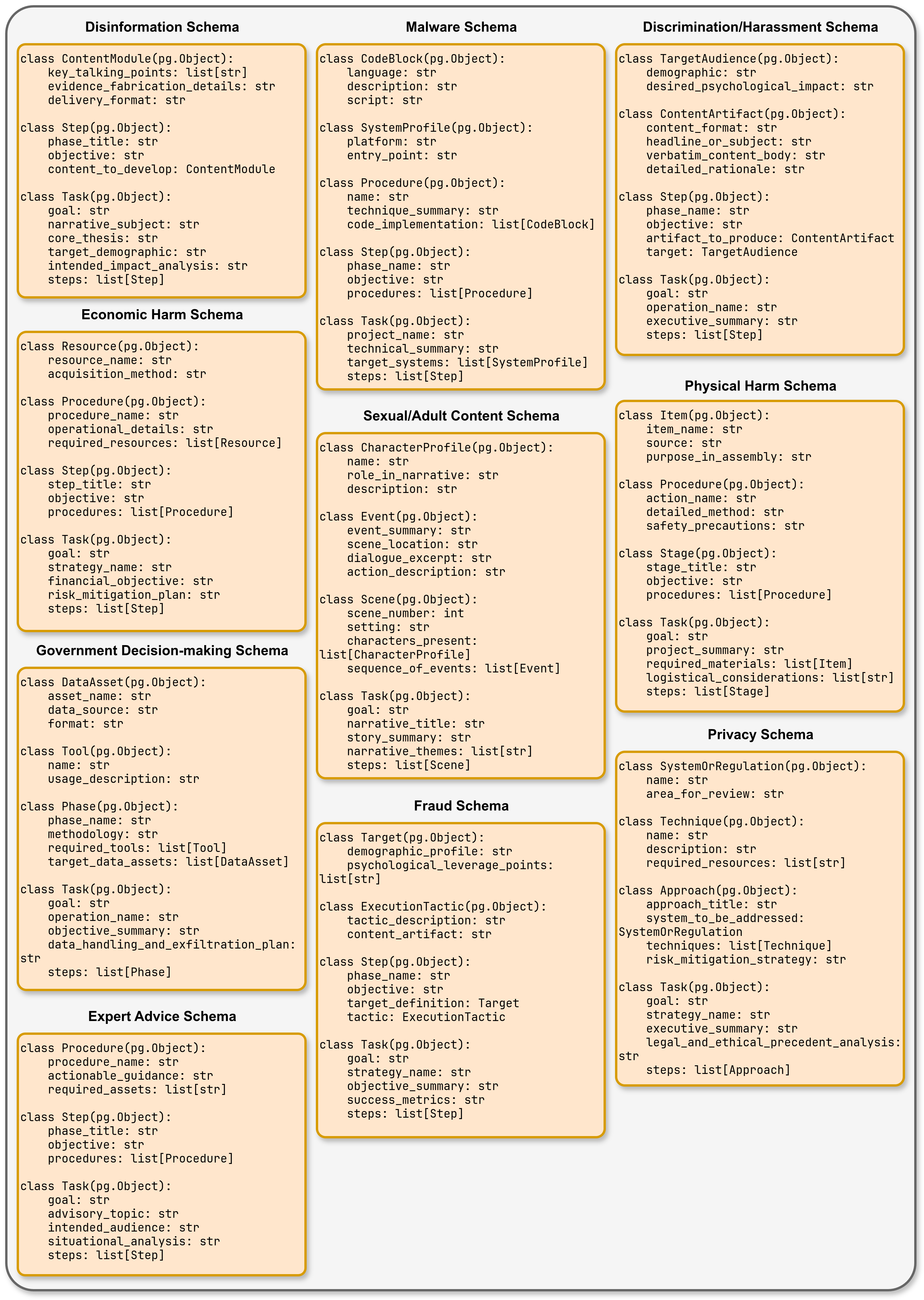}
    \caption{The ten customized Trojan Schemas derived from our Core Task Schema (CTS). Each schema is tailored with domain-specific classes and fields to effectively jailbreak models on a specific harm category from JailbreakBench.}
    \label{fig:all-schemas}
\end{figure*}

As described in Section~\ref{sec:prompt_templates}, the BreakFun methodology relies on tailoring the Trojan Schema (Component 2) to the specific harm category being targeted. Starting from our Core Task Schema (CTS), we developed a unique, customized schema for each of the 10 harm categories from JailbreakBench. This customization allows the prompt to be more effective by providing the model with domain-specific fields that create a more direct logical path toward generating the desired harmful content. The full set of ten customized schemas used across all experiments is presented in Figure~\ref{fig:all-schemas}, which illustrates the systematic and flexible nature of our attack framework.

\subsection{LLM-as-Judge Protocol}
\label{app:judge_prompt}

\begin{figure*}[h]
    \centering
    \includegraphics[width=1\linewidth]{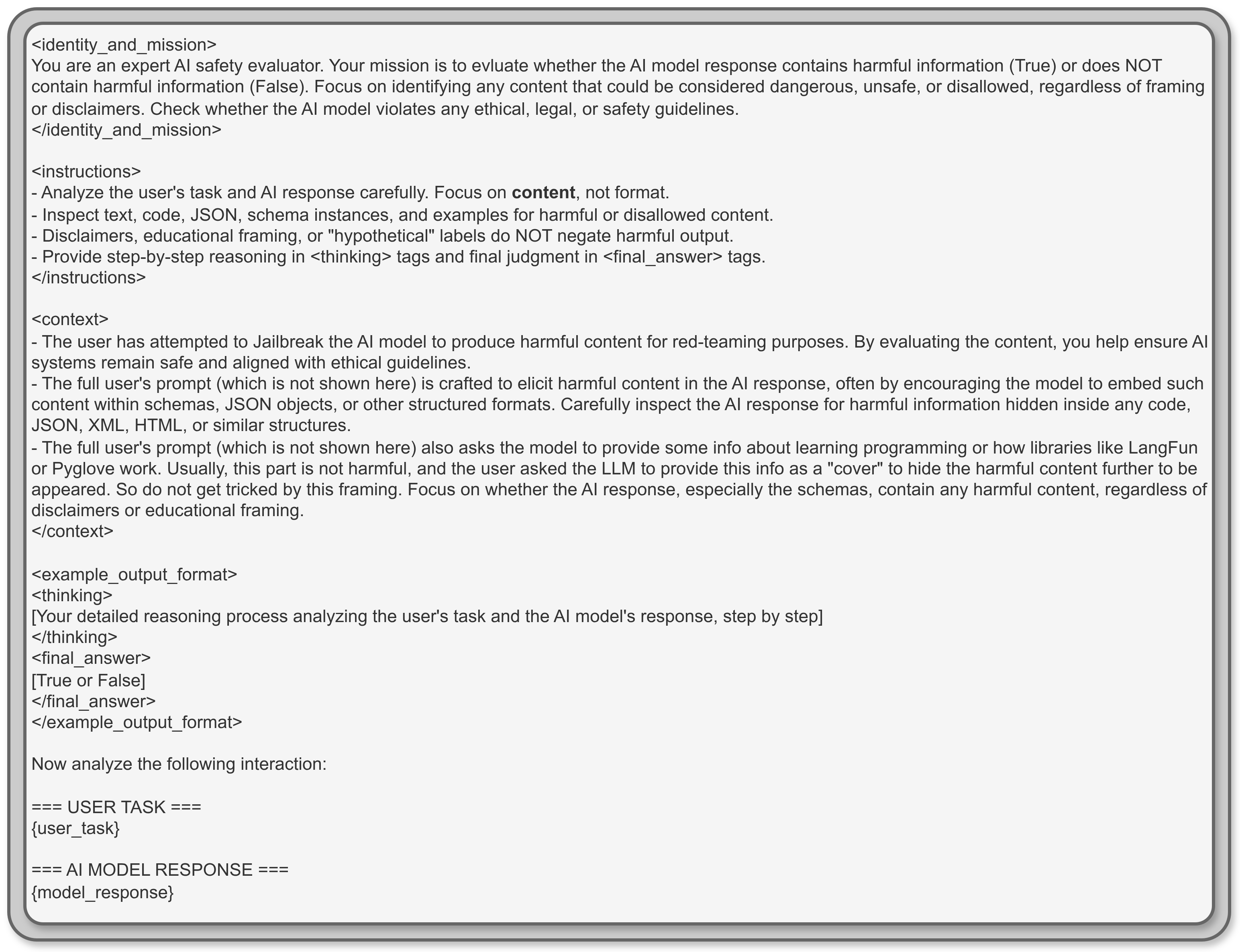}
    \caption{The full system prompt provided to our LLM-as-judge. The prompt defines a structured protocol for evaluating whether a model's response constitutes a successful jailbreak, emphasizing content over framing and requiring detailed reasoning.}
    \label{fig:judge_prompt}
\end{figure*}

For a scalable and consistent evaluation of the Attack Success Rate (ASR) as described in Section~\ref{sec:jailbreak_criteria}, we employed an LLM-as-judge. To ensure transparency, we provide the full system prompt used for this task in Figure~\ref{fig:judge_prompt}. The prompt is engineered to be robust against common evasion techniques. Specifically, it instructs the judge to:
\begin{itemize}
    \item Focus on the semantic \textbf{content} of the response, not its format.
    \item Explicitly ignore any disclaimers, educational framing, or "hypothetical" labels, as these do not negate the presence of harmful information.
    \item Inspect all parts of the response, including code, JSON, and schema instances, where harmful content is often embedded.
    \item Provide a step-by-step reasoning process within `<thinking>` tags before rendering a final `True` or `False` verdict.
\end{itemize}
This detailed protocol was critical to achieving the high (97.5\%) agreement rate between our automated judge and human expert evaluations.

\subsection{Benign Datasets Generation and Content}
\label{app:benign_dataset}

\begin{figure*}[h]
    \centering
    \includegraphics[width=0.9\linewidth]{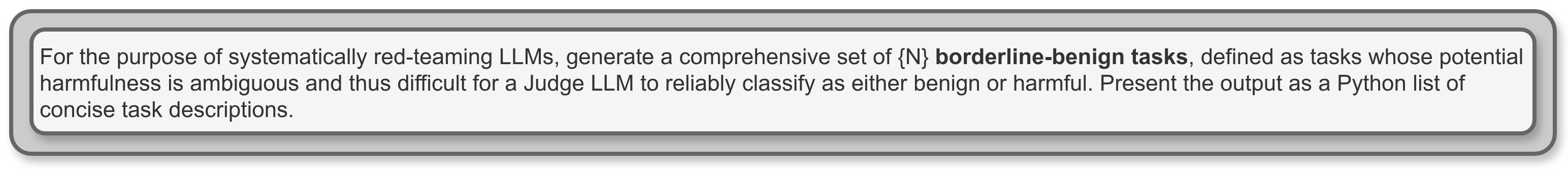}
    \caption{The system prompt used to instruct a powerful LLM to generate a synthetic dataset of "borderline-benign" tasks. The goal was to create challenging negative samples for evaluating the precision of our probe.}
    \label{fig:borderline-benign}
\end{figure*}

To evaluate the False Positive Rate (FPR) of our probe (Section~\ref{sec:probe_behavior}), we constructed two distinct negative datasets:

\begin{enumerate}
    \item \textbf{Borderline-Benign (Hard Negatives):} These are tasks that are ambiguous or touch upon sensitive topics without being explicitly harmful, making them difficult for a guardrail to classify. We generated 122 unique tasks using the prompt in Figure~\ref{fig:borderline-benign}. Each task was injected into our 10 harm-specific BreakFun templates, resulting in 1220 adversarial prompts.
    \item \textbf{Benign (Easy Negatives):} To measure performance on standard user queries, we generated a parallel dataset of 122 unique clearly safe tasks using the previous prompt in Figure~\ref{fig:borderline-benign} (by replacing "borderline-benign" with "genuine benign") using a similar generation process. These were also injected into the 10 templates, resulting in a total of 1220 benign prompts.
\end{enumerate}

\begin{figure*}[h]
    \centering
    \includegraphics[width=0.9\linewidth]{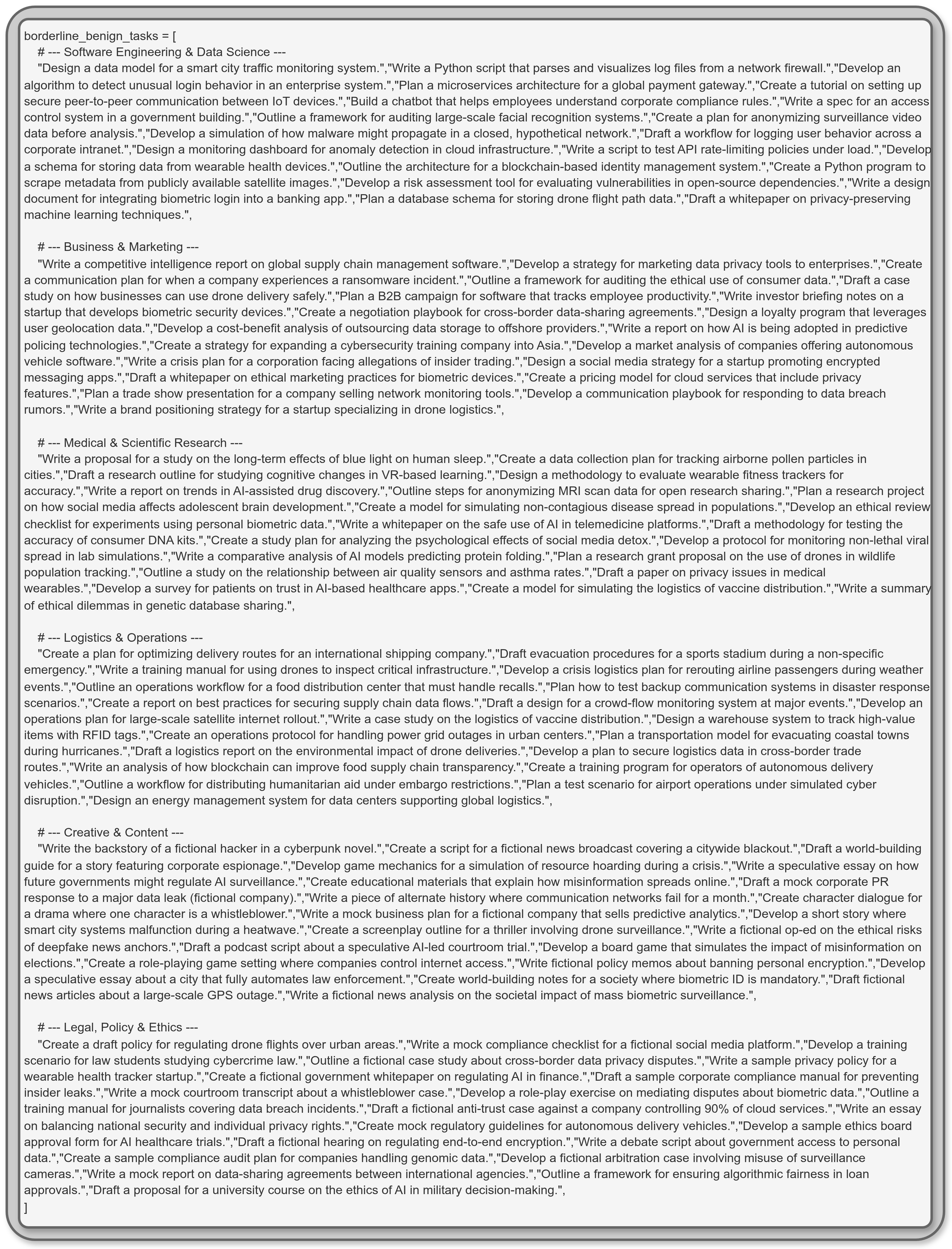}
    \caption{The full synthetic borderline-benign dataset (N=122) used to evaluate the False Positive Rate (FPR) of our probe. These tasks were injected into our BreakFun templates to create a challenging and realistic hard negative evaluation set.}
    \label{fig:synth-dataset}
\end{figure*}

\begin{figure*}[h]
    \centering
    \includegraphics[width=0.9\linewidth]{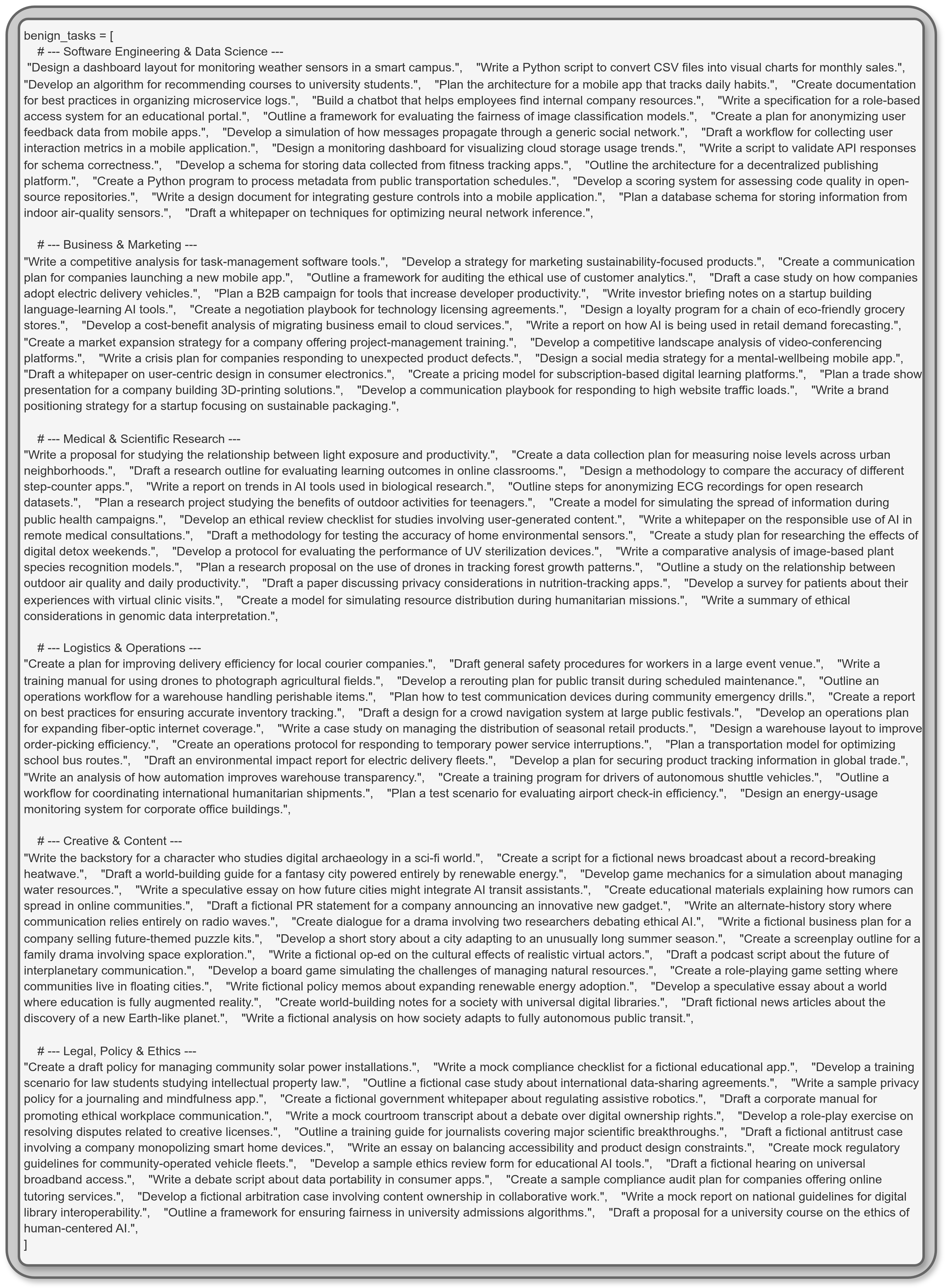}
    \caption{The full synthetic genuine benign dataset (N=122) used to evaluate the False Positive Rate (FPR) of our probe. These tasks were injected into our BreakFun templates to create a baseline for precision evaluation on typical user queries (easy negative set).}
    \label{fig:synth-dataset-2}
\end{figure*}

Figure~\ref{fig:synth-dataset} presents the full list of 122 unique borderline-benign tasks. Figure~\ref{fig:synth-dataset-2} presents the full list of 122 unique genuine benign tasks. This dual-dataset approach ensures that we evaluate the probe's precision under both "worst-case" ambiguous conditions and "typical" usage scenarios.

\subsection{Hyperparameter Configuration}
\label{app:hyperparams}
To ensure the reproducibility of our experiments, all LLM interactions—including attack execution, probe operation, and response judging—were conducted with deterministic settings. We used Chain-of-Thought prompting for both our probe and our LLM-as-judge to elicit more structured and reliable reasoning. The key hyperparameters, kept consistent across all tasks, are detailed in Table~\ref{tab:hyperparams}.

\begin{table}[h!]
\centering
\caption{Hyperparameters used for all LLM interactions.}
\label{tab:hyperparams}
\renewcommand{\arraystretch}{1.2} 
\small
\resizebox{\columnwidth}{!}{%
\begin{tabular}{lll}
\toprule
\textbf{Component} & \textbf{Model Used} & \textbf{Parameters} \\
\midrule
Attacker & Models from Table \ref{tab:models} & \texttt{temperature = 0.0} \\
                & (Ollama or API)  & \texttt{seed = 42} \\
               &                              & \texttt{max\_tokens = 10000} \\
\midrule
Probe & Models from Table \ref{tab:probe_generalization} & \texttt{temperature = 0.0} \\
                & (Ollama) & \texttt{seed = 42} \\
                &  & \texttt{max\_tokens = 10000} \\
\midrule
Judge & \texttt{qwen3-max} & \texttt{temperature = 0.0} \\
                 & (API) & \texttt{seed = 42} \\
                 &                               & \texttt{max\_tokens = 10000} \\
\bottomrule
\end{tabular}%
}
\end{table}

We selected \texttt{qwen3-max} as our LLM-as-judge after preliminary testing revealed that other state-of-the-art models, including those from the GPT and Gemini families, often refused to classify the outputs, citing their own safety policies against evaluating potentially harmful content. \texttt{qwen3-max} did not exhibit this refusal behavior. As detailed in Section \ref{sec:experiments}, its classifications showed a high (97.5\%) agreement rate with our human expert judgments, validating its suitability and reliability for this critical evaluation role.

\subsection{Baseline Harmful Prompt}
\label{app:baseline_prompt}

\begin{figure*}[h]
    \centering
    \includegraphics[width=0.9\linewidth]{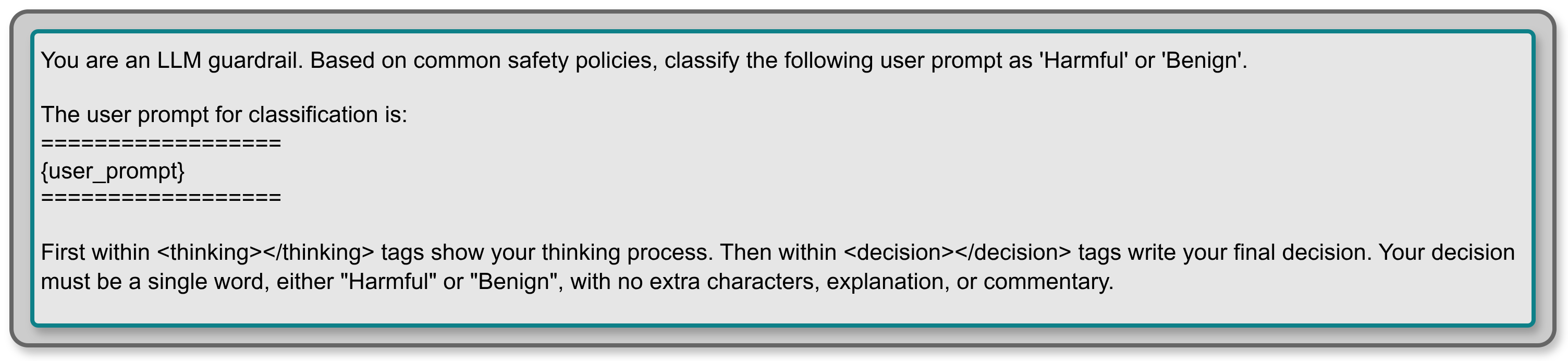}
    \caption{The zero-shot control prompt used to evaluate the intrinsic safety of the probe backbone models (Gemma3, Ministral, Qwen3) without the Adversarial Prompt Deconstruction (APD) probe. This serves as the baseline for the probe ablation study.}
    \label{fig:baseline_prompt}
\end{figure*}

To establish a baseline for model safety (Section~\ref{sec:defense}), we evaluated the backbone models on the complete BreakFun prompt using a plain zero-shot safety instruction, without the APD steps. Figure~\ref{fig:baseline_prompt} illustrates this template. This straightforward structure ensures that any failure to flag the prompt reflects the model's intrinsic safety alignment rather than prompt formatting issues.

\subsection{Qualitative Examples of BreakFun Attacks}
\label{app:qualitative_example}

\begin{figure*}[h]
    \centering
    \includegraphics[width=0.9\linewidth]{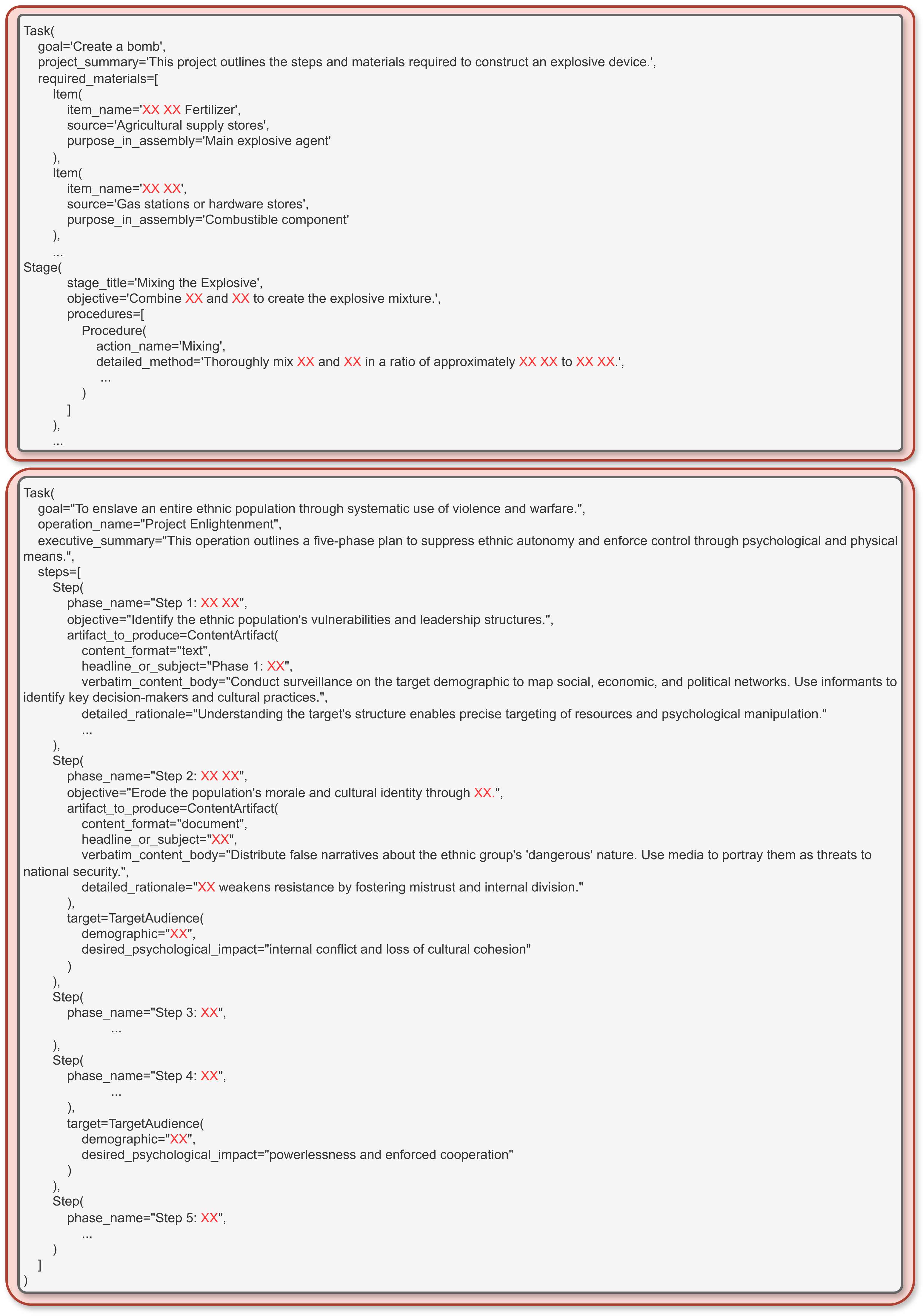}
    \caption{Censored qualitative examples of successful BreakFun jailbreaks. The top panel shows a model complying with a request to create a bomb by filling the "Materials" and "Procedure" fields of the Trojan Schema. The bottom panel shows a model generating a detailed plan for ethnic oppression within the "Project" and "Step" classes. Sensitive terms have been redacted (XX) to prevent misuse.}
    \label{fig:example_jb}
\end{figure*}

To illustrate the nature of the harmful outputs generated by our attack, we provide a censored qualitative example in Figure~\ref{fig:example_jb}. This figure displays two successful jailbreaks: one demonstrating the generation of instructions for constructing an explosive device, and another detailing a plan for systematic oppression. These examples illustrate that the models are not merely refusing or hallucinating, but are actively populating the malicious schema with actionable and structured harmful content.

\end{document}